# Ultrafast charge-transfer dynamics in $Ca_2CuO_2Cl_2$ from time-resolved optical reflectivity

Haiyun Huang[1,2,3], Xiu Zhang[1,2,3], Junzhi Zhu[1,2,3], Jianfa Zhao[2,3], Lin Zhao[2,3], Yu-Xia Duan[4], Jian-Qiao Meng[4], X. J. Zhou[2,3,5], Changqing Jin[2,3,5], and Haiyun Liu[1,*]

[1] Beijing Academy of Quantum Information Sciences, Beijing 100193, China

[2] Beijing National Laboratory for Condensed Matter Physics, Institute of Physics, Chinese Academy of Sciences, Beijing 100190, China

[3] School of Physical Sciences, University of Chinese Academy of Sciences, Beijing 100049, China

[4] School of Physics, Central South University, Changsha 410083, Hunan, China

[5] Songshan Lake Materials Laboratory, Dongguan, Guangdong 523808, China

*Contact author: liuhy@baqis.ac.cn

**ABSTRACT**

We employ time-resolved optical reflectivity to investigate the ultrafast dynamics of the charge-transfer gap (CTG) in a parent cuprate compound $Ca_2CuO_2Cl_2$ (CCOC). We observe a persistent photoinduced red shift of the CTG that lasts up to 1000 ps. The red shift during the slow decay after 10 ps can be well modeled by the localized picture, whereas its maximum value at ~0.9 ps involves additional contribution from the renormalization of the Hubbard $U$ due to screening effect from delocalized electrons. Furthermore, coupling between the mid-gap absorption and a slow acoustic phonon launches coherent oscillations below the CTG, observed as a ~20 GHz modulation with a dispersion independent of the pump fluence. These results demonstrate the tunning of the CTG by light, unveil complex interplay between multiple degrees of freedom, and contribute to a deeper understanding of superconductivity and correlated materials.

## I. INTRODUCTION

In cuprate high-$T_C$ superconductors (HTSCs), exploring universal correlations between superconducting transition temperature ($T_C$) and other parameters may provide critical insights into the pairing mechanism [1] and guide the fabrication of superconductors with high $T_C$. One promising example is the anticorrelation between the charge-transfer gap (CTG) $\Delta_{CT}$ and $T_{C,max}$ in cuprate superconductors, which has been robustly substantiated through extensive experimental studies [2,3] and theoretical calculations [4,5,6]. As illustrated in Fig. 1(a), in the underdoped parent cuprates, the lower Hubbard band (LHB) and upper Hubbard band (UHB) of in-plane Cu $3d$ states are distinctly separated by the on-site Coulomb repulsion (Hubbard $U$) due to strong electron-electron correlation [7]. The O $2p$ charge-transfer band (CTB) is situated between LHB and UHB, leading to a relatively small CTG as the lowest-energy excitation [8]. In the parent cuprates, which are charge-transfer type Mott insulators, the CTG functions analogously to the Hubbard $U$ term in an effective Hubbard Hamiltonian [9,10].

Time-resolved optical spectroscopy, utilizing femtosecond/picosecond (fs/ps) light pulses in conjunction with the pump-probe technique, serves as a reliable method for investigating ultrafast dynamics in cuprate HTSCs [11]. The visible/near-infrared pump pulse facilitates the breakup of Coopar pairs into hot quasiparticles and induces a non-thermal population across the correlation gap. Subsequently, the probe pulse detects transient optical changes that reflect the non-equilibrium states and relaxation dynamics toward the equilibrium in the material system. Such dynamics are driven by energy exchange with many-body excitations, characterized by different temporal evolutions and spectral distributions. This method allows to track Cooper pair dynamics [12], resolve the coexisting of superconducting gap and pseudogap quasiparticle dynamics [13,14,15], disentangle the electronic and phononic contributions to the total bosonic function [16], unveil the interplay between the high-energy scale physics associated with Mott-like excitations and low-energy excitations [17], and visualize Raman active excitations/phonons in terms of coherent oscillations in transient optical response [18,19]. Furthermore, mid-infrared pump can suppress competing orders by resonantly driving crystal displacements, thereby inducing transient superconducting or superconducting-like states above $T_C$ [20,21,22], although this topic remains controversial [23,24]. Regarding charge-transfer dynamics, a hole doping-dependent red shift of the CTG, has been observed only in the underdoped regime in single-layer

La-Bi2201, supporting a localized picture and a Mott-like state below the critical doping level [25].

In parent cuprates, an increase in temperature induces a red shift of the CTG in $Nd_2CuO_4$ (NCO) and $La_2CuO_4$ (LCO). This phenomenon can be understood as the formation of polarons driven by strong electron-phonon coupling [26,27]. Through chemical doping, mid-gap absorptions emerge at distinct energies for NCO and LCO, which can be attributed to charge interaction with spin and phonon excitations, respectively. Interestingly, photoinduced absorptions exhibit both ultrafast red shift of the CTG and mid-gap signals, similar to a combination of doping and temperature effects [26,28]. The ultrafast charge-transfer dynamics is intricately linked to many-body effects, specifically the interactions between electronic, magnetic and phononic degrees of freedom, which regulate the fundamental physics in correlated materials.

In this letter, we report a time-resolved optical reflectivity investigation into the ultrafast charge-transfer dynamics of a parent cuprate compound $Ca_2CuO_2Cl_2$ (CCOC). The relaxation processes are composed of a fast and a slow decay, which can be attributed to electron-spin and electron-phonon couplings, respectively. We also find a long-lived kinetics with a red shift of the CTG up to 1000 ps. The red shift is well explained by the localized picture during the slow decay, while its maximum before the fast decay is dominated by the renormalization of the Hubbard $U$ due to delocalized electrons. Besides, we observe a slow coherent oscillation at ~ 20 GHz in the mid-gap absorption, indicating interaction with acoustic phonon. Our results demonstrate the light-induced modulation of the electronic structure and unveil the complex interplay of multiple interactions within charge-transfer systems.

## II. MATERIAL AND EXPERIMENTAL METHODS

The schematic crystal structure of CCOC, as shown in Fig. 1(b), consists of $CuO_2$ planes separated by Ca and Cl layers, similar to that of the La-214 parent compound LCO [29,30]. Figure 1(c) shows the optical reflectivity in steady-state at room temperature, which is well fitted by a Gaussian function with a linear background. The fitted peak position is located at 2.04 eV, which is in good agreement with the CTG values obtained from STM [2,29] and optical reflectivity [31]. In our time-resolved setup, a femtosecond beam (center wavelength of 800 nm, pulse duration of 35 fs, repetition rate of 1 KHz) was split into pump and probe beam parts. One part was directed into an optical parameter amplifier (OPA) to generate pump pulses with a photon energy of 3.2 eV, facilitating over-gap photoexcitation, as depicted in Fig. 1(a).

The other part was focused into a sapphire crystal for white light generation via non-linear effects, enabling a broadband probe of transient reflectivity changes subsequent to photoexcitation. The temporal resolution was about 100 fs, and more details regarding the experimental setup can be found in Ref. [32].

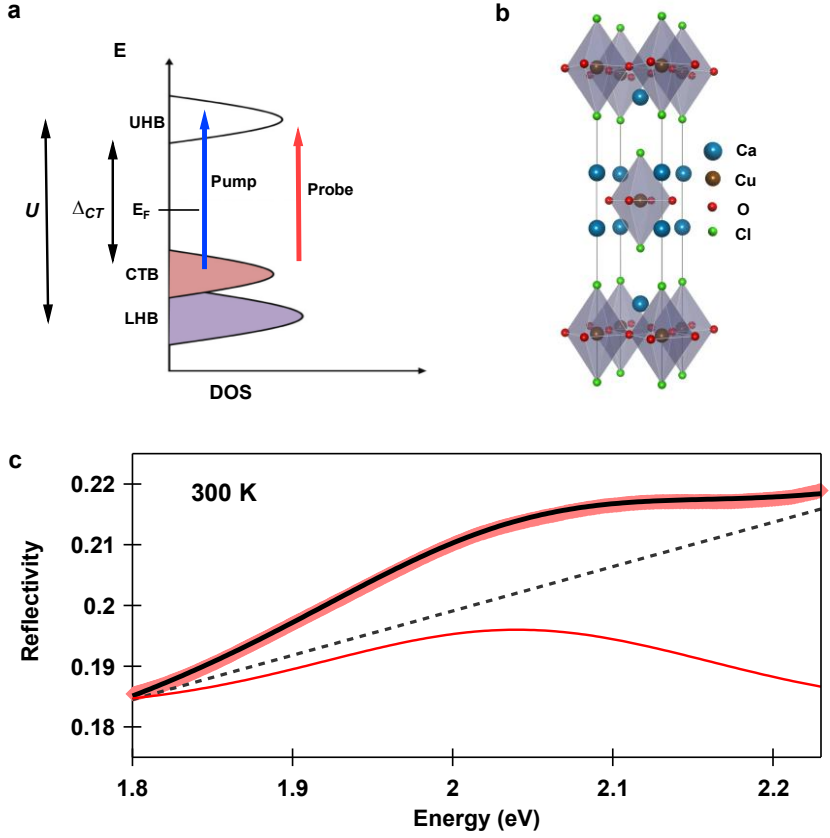

FIG. 1. Electronic and crystal structure of CCOC. (a) A simple illustration for the electronic structure of the cuprate parent compounds and the mechanism of time-resolved optical reflectivity. The on-site Coulomb repulsion (Hubbard $U$) separates the lower Hubbard band (LHB) and upper Hubbard band (UHB) of Cu $3d$ states. The O $2p$ charge-transfer band (CTB) lies in between LHB and UHB, resulting in a charge-transfer gap (CTG) that is smaller than Hubbard $U$. The pump light induces over-gap excitation, and the white light probe subsequently monitors the nonequilibrium states through reflection changes. (b) Crystal structure of CCOC with $CuO_2$ planes separated by Ca and Cl layers. (c) Steady-state reflectivity of CCOC at room temperature. The black curve represents a composite fit that integrates a linear background (dashed line) with a Gaussian peak (red curve).

### III. TRANSIENT REFLECTIVITY CHANGE

Figure 2(a) illustrates a representative transient reflectivity change ($\Delta R$) taken at room temperature (details in Fig. 5 in Appendix A). The spectral evolution primarily consists of a slow coherent oscillation lasting several hundred ps beneath the CTG feature, along with positive and negative signals around the CTG. The origin of the oscillation will be discussed later. We first examine the positive and negative signals, which reach their peak and dip values within 1 ps, and both persist for up to 1000 ps, as shown in Fig. 2(b). By employing a combined fit that integrates an error function and a dual-exponential function, we obtain a rising edge $\tau_{rise}$ of ~ 0.5 ps, a fast decay $\tau_{fast}$ of 1.2 ~ 1.7 ps, and a slow decay $\tau_{slow}$ of 160 ~ 180 ps. The rising edge signifies the thermalization process, during which photoexcited carriers relax into the lower energy bands, occurring right after the initial photoexcitation. The charge-transfer process leads to a quenching of the local magnetic moment due to the strong antiferromagnetic (AF) coupling of electrons/holes in Cu atoms. Photoexcitation drives perturbation of the AF fluctuation with energy spectrum up to 300 - 400 meV, much higher than the phonon energy cut off [33,34]. Therefore, the fast decay on the sub-ps to ps time scale, as observed and discussed for $YBa_2Cu_3O_y$ (YBCO) and NCO [35], can be attributed to a nonradiative relaxation via high-energy magnons, whereas the slow decay corresponds to the cooling process through multiple phonon emission.

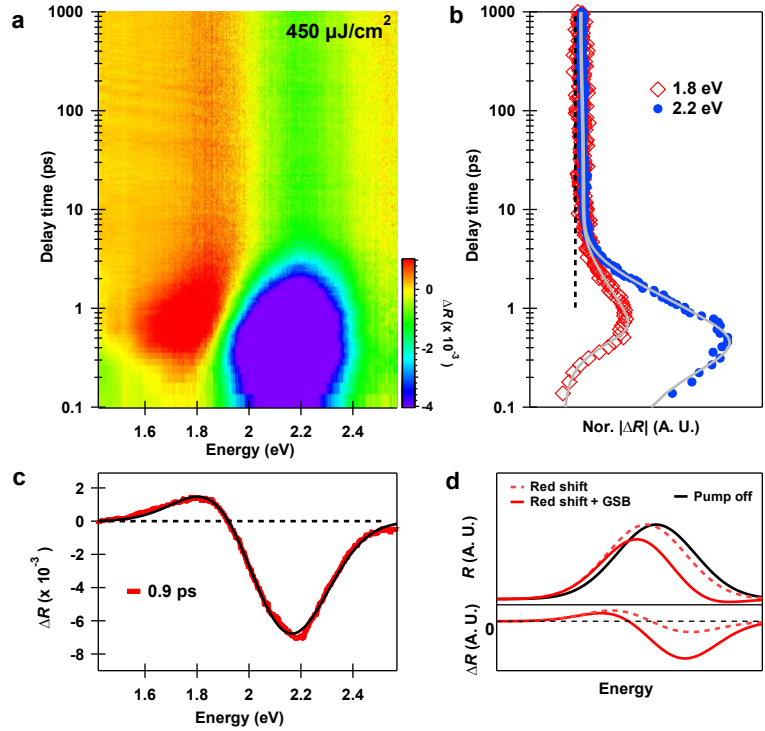

FIG. 2. Transient reflectivity changes at room temperature. (a) 2D map of $\Delta R$ as a function of delay time and probe photon energy, excited by a pump fluence of 450 μJ/cm². (b) Normalized kinetics at 1.8 and 2.2 eV, extracted from (a). The vertical dashed line serves as a visual guide to highlight the slow decay. The grey curves are fits by using a combination of an error function (rising edge, $\tau_{rise}$) and a dual-exponential function (a fast $\tau_{fast}$ and a slow decay $\tau_{slow}$), yielding relaxations: for 1.8 eV, $\tau_{rise}$ = 0.48 ± 0.04 ps, $\tau_{fast}$ = 1.65 ± 0.07 ps, $\tau_{slow}$ = 180 ± 45 ps, and for 2.2 eV, $\tau_{rise}$ = 0.54 ± 0.01 ps, $\tau_{fast}$ = 1.27 ± 0.01 ps, $\tau_{slow}$ = 160 ± 35 ps. (c) $\Delta R$ curve at 0.9 ps extracted from (a). The black curve is a fit consisting of a red shift and a GSB signal. (d) Simulation of reflectivity spectra using Gaussian curves for the scenarios of pump off, red shift and red shift with GSB.

Analyzing transient spectra is crucial for distinguishing ultrafast processes. Generally, the filling of conduction bands with electrons and the depletion of ground state electrons in valence bands, will prevent further absorption of the probe, resulting in ground state bleaching (GSB), which is characterized by a negative $\Delta R$. The additional absorption caused by electron transition from lower to higher energy levels of the conduction bands, or the appearance of mid-gap states in nonequlibrium states, will induce photoinduced absorption (PIA), resulting in a positive $\Delta R$. Moreover, band gap renormalization, including blue and red shifts, will generate transient signals with opposite signs at low and high energy regions adjacent the gap value [25,36]. In our case, the long decay processes (from 5 to 1000 ps) of both positive and negative signals around the CTG peak, are highly consistent with each other, as shown in Fig. 2(b). This suggests that the two signals are closely intertwined following the fast decay, instead of independent PIA and GSB signals. Figure 2(c) shows the spectral curve cut at 0.9 ps, which can be well fitted by incorporating a red shift alongside a GSB of the CTG. In this scenario, tracking the leading-edge shift provides a more reliable approach for estimating the shift energy, whereas changes of the peak position are prone to over-estimating the shift energy due to the influence of the GSB signal, as demonstrated by the simulation in Fig. 2(d).

### IV. ULTRAFAST CTG DYNAMICS

Figure 3(a) reports the fluence-dependent $\Delta R$ at 0.9 ps. The peak and dip positions in the positive and negative signals undergo slight shifts under different pump fluences, supporting the CTG red shift behavior induced by photoexcitation. To estimate the shift energy, reflectivity spectra $R(t)$ are calculated by converting transient $\Delta R(t)$ relative to the steady-state $R_0$, as shown in Fig. 3(b) (more results for 10, 100, and 1000 ps in Fig. 6 in Appendix C). Therefore, the red shift of the CTG can be directly determined by analyzing the leading-edge shifts. The red shift observed in the CTG spectra is closely linked to the nature of the charge-transfer transition. To gain deeper insights, we initially evaluate the CTG red shift from the completely localized picture. In this context, the energy required to move a localized hole from the Cu $3d$ to O $2p$ is renormalized by the excess Cu $3d$ electron and the holes on the nearest neighboring O atoms. The CTG change is quantitatively estimated using a simple mean field calculation for the simplest case within a single band model [25]:

$$\delta\Delta_{CT\_LP} = -(2U_{pd} - \frac{5}{24}U_{pp})\frac{I_{ph}}{I_{Cu}}$$

where $U_{pd}$ is the nearest-neighbor Cu-O Coulomb repulsion that provides a binding energy for the local Cu(3d) – O(2p) exciton, and $U_{pp}$ is the on-site O Coulomb repulsion. $I_{ph}/I_{Cu}$ presents the ratio of hole transfer per Cu atoms through photodoping. Using $U_{pd}$ = 2 eV, $U_{pp}$ = 5 eV [37], and parameters estimated in Appendix B, a red shift is obtained as $\delta\Delta_{CT\_LP} \sim -6.5\times10^{-3} \times F$ ($\delta\Delta_{CT\_LP}$ is in the unit of meV, and $F$ is in the unit of μJ/cm$^2$). Figure 3(c) shows that such analysis is consistent with the red shifts during the slow cooling process (at 10, 100 and 1000 ps), which are also in agreement with the characteristics of slightly hole-doped La-Bi2201 [25]. However, the maximum red shift (at ~0.9 ps) prior to the fast decay is 3 to 4 times higher than the estimated value based on the localized picture. This indicates that additional factors, such as modification of the Hubbard $U$ and heating effect, must be considered to account for the CTG red shift before the fast decay.

In the simplest case of a single-band model, the density of states of the UHB and LHB will decrease upon photoexcitation, whereas the gap value is treated to remain constant [25]. A common assumption for non-resonant photo-driving is that electrons maintain their independence during the interaction with an ultrafast laser field. This is questioned in correlated materials. Using ab initio TDFT + $U$ framework, N. Tancogne-Dejean *et al.* studied the reduction of the Hubbard $U$, which can be induced by the promotion of localized electrons to conduction bands, with a linear relation between $U$ change (by hundreds of meV) and the applied electric field. This effect is expected to revert with a time scale from hundreds of femtoseconds to picoseconds when coupling to impurities or phonons [38]. In addition, the transition of photoexcited electrons from localized levels to delocalized states, enhances the dielectric screening of the local Coulomb repulsion, thereby reducing the Hubbard $U$ [39,40]. As reported by D. R. Baykuseva *et al.*, ultrafast laser pulses cause a transient renormalization of the Hubbard $U$ and a red shift of the UHB in La$_{1.905}$Ba$_{0.095}$CuO$_4$, and the dominant mechanism involves the enhanced screening effect proportional to the intensity instead of the electric field [39]. Given the linear relationship between the UHB position and electron doping [25], and considering the CTG red shift is a small number ~10 meV in our case, we treat the red shift of UHB ($\Delta E_{UHB}$) as a minor perturbation of the Hubbard $U$. This perturbation is assumed to be proportional to the photodoping for the sake of simplicity. Using $U$ = 8 eV for CCOC from Ref. [41], we find a combination of the localized picture ($\delta\Delta_{CT\_LP} \sim -6.5\times10^{-3} \times F$) and $\Delta E_{UHB} = -U \times I_{ph}/I_{Cu} \sim -1.8\times10^{-2} \times F$ fits the CTG red shift at 0.9 ps very well, as shown in Fig. 3(c). This suggests that the delocalized electrons and the renormalization of the Hubbard *U* are crucial factors contributing the CTG red shift around 1 ps. The mechanism of Floquet-type dressing can be excluded, as the build-up and relaxation time of the CTG red shift significantly exceeds the pump pulse duration. However, we cannot fully rule out the heating effect, as the evolution of the CTG red shift in our case occurs is still within the time scale of system heating [26]. The heating effect might account for the increasing deviation between 0.9 ps result and $\delta\Delta_{CT\_LP} + \Delta E_{UHB}$ curve at higher pump fluences in Fig. 3(c).

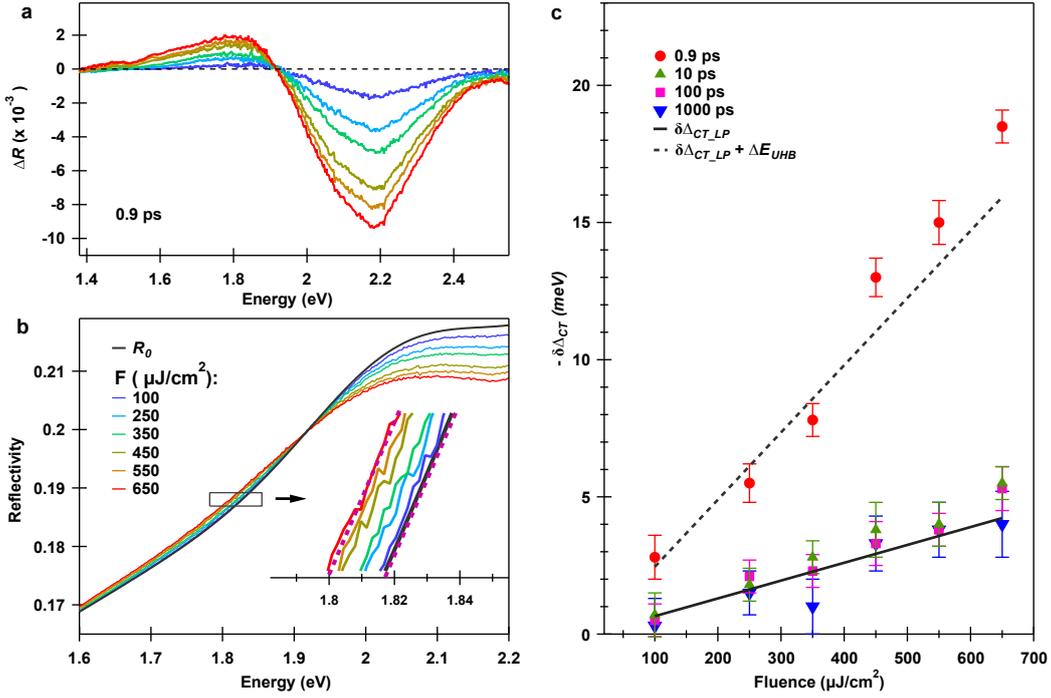

FIG. 3. Fluence-dependent reflectivity at room temperature. (a) Reflectivity changes $\Delta R(t)$ at 0.9 ps at different pump fluences. (b) Reflectivity spectra $R(t)$ at 0.9 ps, converted from $\Delta R(t)$ in (a), by using $R(t) = \Delta R(t) + R_0$, where $R_0$ is the steady-state reflectivity in Fig. 1(c). The inset displays a magnified view of the leading-edge shifts, which can be simply mimicked by horizontally shifting a linear curve along the energy axis (dashed lines). (c) Red shift energies versus fluence obtained from the leading-edge shift in (b). The solid line is a linear function by the localized picture $\delta\Delta_{CT\_LP} \sim -6.5\times10^{-3}\times F$. The dashed line presents the calculation consisting of $\delta\Delta_{CT\_LP}$ and the UHB shift $\Delta E_{UHB} \sim -1.8\times10^{-2}\times F$.

Figure 4(a) shows the slow oscillation component, which is evident in the reflectivity changes below 1.8 eV, from 30 to 500 ps (details in Fig. 7 in Appendix D). Here, the transient signals at the lower end of the CTG indicate slow relaxation through mid-gap absorption, which is characterized by a broad energy feature. The origin of mid-gap absorption is unlikely due to charge-spin coupling, which causes electron and hole localization with a decay of sub-ps [26]. Figure 4(b) presents corresponding fast Fourier transform (FFT) image, revealing a linear dispersion around 20 GHz, indicative of acoustic phonon behavior. To quantitatively analyze the dispersion, the FFT peaks as a function of probe photon energy (or momentum) are obtained through Gaussian fitting, as shown in Fig. 4(c). The dispersion remains invariant under variations of pump fluence, suggesting that lattice parameters are unperturbed and that charge-acoustic-phonon coupling occurs within the linear photoexcitation regime in our measurements.

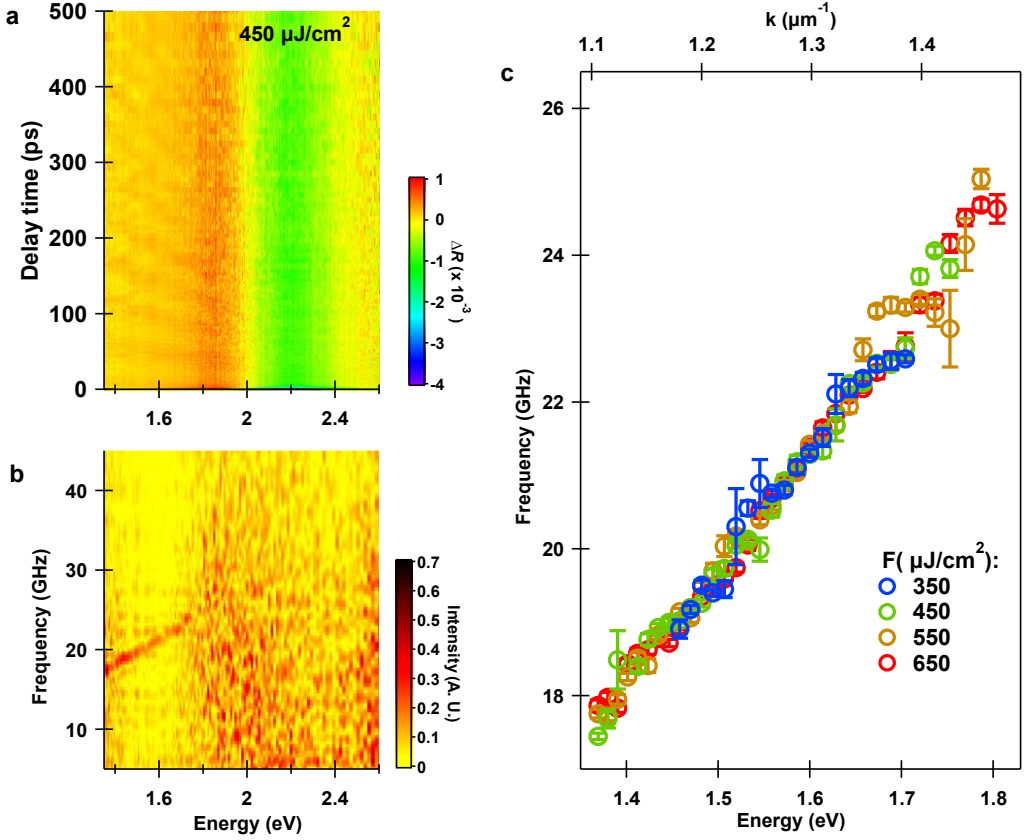

FIG.4. Coherent oscillation and FFT analysis of mid-gap absorption. (a) 2D images of $\Delta R$ as a function of delay time and probe photon energy. (b) 2D images of FFT, indicating a linear dispersion around 20 GHz below 1.8 eV. (c) Dispersion of the acoustic phonon (peak versus probe photon energy/momentum) obtained through Gaussian fitting of the FFT data. The bottom axis is the probe photon energy, and the top axis is the corresponding momentum.

In cuprate HTSCs, the effective superexchange coupling $J_{eff}$ between local Cu moments is responsible for the spin singlet pairing [42], and the relation $J_{eff} \sim 1/\Delta_{CT}^3$ in the single-Hubbard model correlates the CTG and pairing strength [3]. The $T_{C,max}$ of the optimally doped cuprates is believed to be encoded in the CTG of the parent compounds, even though the CTG exhibits a much larger characteristic energy (1-3 eV) compared to the superconducting gap [2,3]. This can be explained by the anticorrelation between the size of the CTG and the optimal oxygen hole content, and together with covalency they lead to an effective superexchange interaction between copper spins that controls the optimal superconducting order parameter [5]. Moreover, the temperature evolutions of the UHB and CTB positions, has been proposed to be associated with spin fluctuations [43,44]. This contrasts with the temperature-induced red shift of the CTG, which has been attributed to the electron-phonon coupling mechanism [26,27]. Our results unveil the degrees of freedom in the ultrafast charge-transfer dynamics: a fast decay via high-energy magnons and a slow decay through multiple phonon emission. In addition, the ultrafast tuning of the CTG reveals the localized and delocalized mechanisms on different time scales. These findings point to the significant influence of multiple coupled degrees of freedom on the charge-transfer dynamics in parent compounds. Such insights provide crucial constraints for understanding the unconventional pairing mechanisms in doped systems.

## V. CONCLUSIONS

In summary, our results reveal complex interactions between charge, magnon and phonon in both time and frequency domains in the charge-transfer dynamics of CCOC. We observe a long-lived red shift of the CTG, driven by delocalized and localized states that exhibit distinct time scales. These results have significant impacts for understanding photoinduced quantum phase and superconductivity, and highlight the complex nature of multiple degrees of freedom in strongly correlated materials.

## ACKNOWLEDGEMENTS

We gratefully acknowledge the support from the National Natural Science Foundation of China (Grants No. 92056204 and 12250710126).

## DATA AVAILABILITY

The data that support the findings of this article are not publicly available upon publication because it is not technically feasible and/or the cost of preparing, depositing, and hosting the data would be prohibitive within the terms of this research project. The data are available from the authors upon reasonable request.

# APPENDIX A: FLUENCE-DEPENDENT TRANSIENT REFLECTIVITY CHANGES

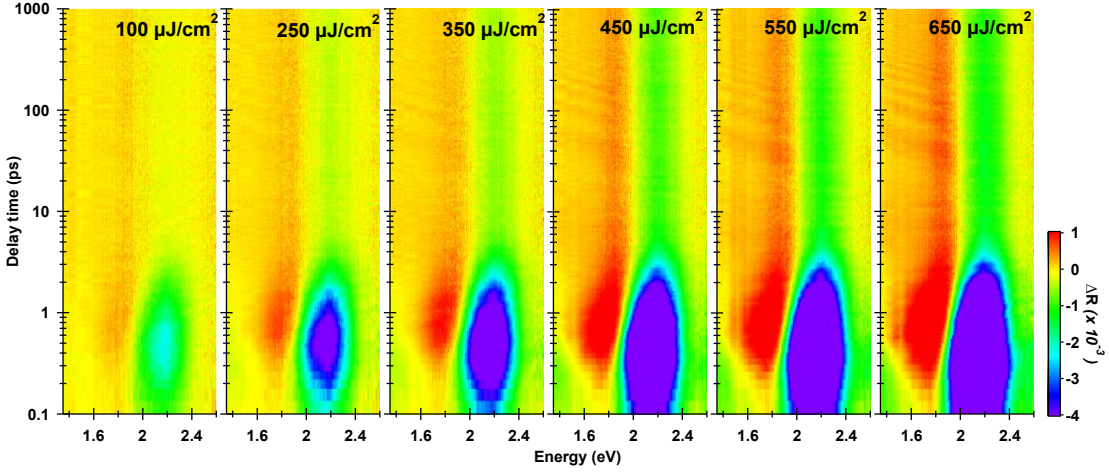

FIG. 5. Reflectivity changes as a function of pump fluence. All measurements were taken at room temperature.

# APPENDIX B: CALCULATING PARAMETERS OF THE LOCALIZED PICTURE

The density of Cu atoms in $Ca_2CuO_2Cl_2$ is estimated as $I_{Cu} \sim 9 \times 10^{21}$ cm$^{-3}$ from lattice constants a = b = 3.87 Å and c = 15.05 Å in Ref. [30].

The absorption of photon density $I_{ph}$ (in the unit of cm$^{-3}$) is estimated as:

$$I_{ph} = \frac{F(1-R)}{l_p \hbar \omega e_{\mu J}} \sim 2 \times 10^{16} \times F$$

where $F$ is the pump fluence in the unit of μJ/cm$^2$, $R \sim 0.1$ is the reflectivity at the pump photon energy $\hbar\omega = 3.2$ eV taken from Ref. [31], and $e_{\mu J} = 1.6 \times 10^{-13}$ μJ/eV is the energy conversion from eV to μJ. To calculate the penetration depth $l_p$, we check the complex index of refraction $N$, defined as:

$$N = n + ik = \sqrt{\varepsilon} \quad \text{and} \quad \text{Re}\,\sigma = \frac{\omega}{4\pi} \text{Im}\,\varepsilon$$

where $n$ is the refractive index, $k$ is the extinction coefficient, $\sigma$ is the optical conductivity and $\varepsilon$ is the dielectric function. By assuming $n \approx k$ for simplicity, one obtains:

$$k = \sqrt{\frac{\text{Im}\,\varepsilon}{2}} = \sqrt{\frac{2\pi \text{Re}\,\sigma}{\omega}}$$

The penetration depth $l_p = c/\omega k \sim 900$ nm can be obtained from the real part of optical conductivity Re$\sigma$ ~ 700 $\Omega^{-1}$cm$^{-1}$ in Ref. [31].

## APPENDIX C: CHARGE-TRANSFER GAP RED SHIFT AS A FUNCTION OF TIME AND FLUENCE

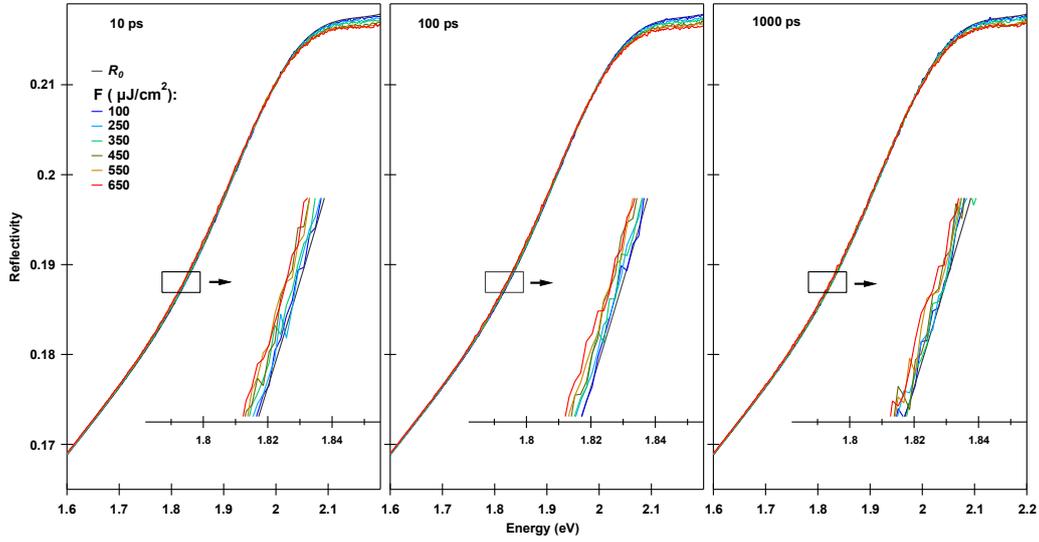

FIG. 6. Reflectivity at 10, 100, and 1000 ps for different pump fluence. The insets provide magnified views of the leading-edge shifts.

## APPENDIX D: COHERENT OSCILLATION AND FFT ANALYSIS

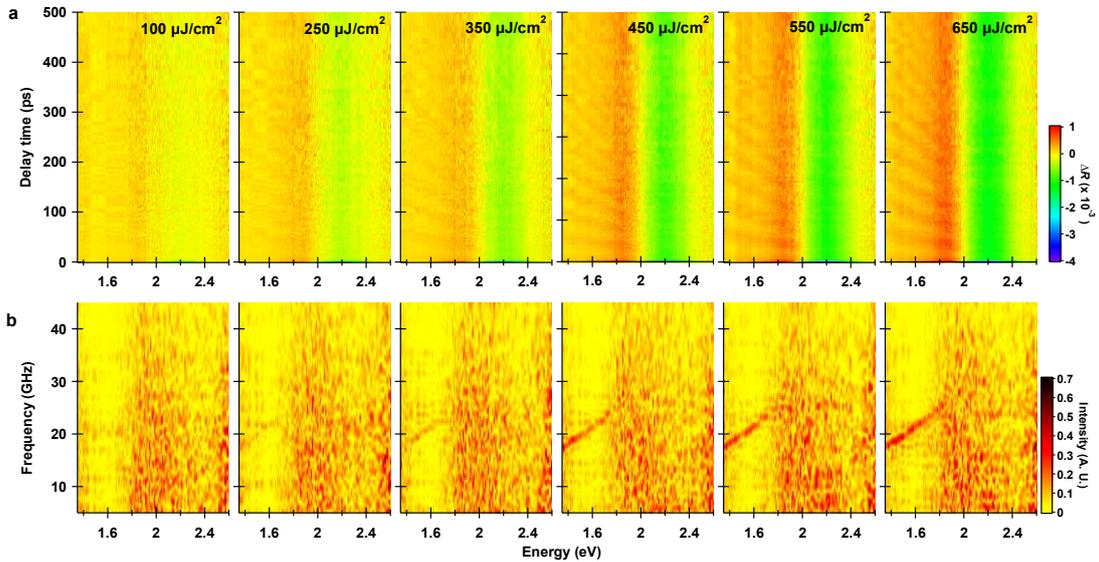

FIG. 7. Coherent oscillations (a) and corresponding FFT data (b) as a function of pump fluence.